\documentclass[%
 reprint,
superscriptaddress,
 amsmath,amssymb,
 aps,
]{revtex4-2}

\usepackage{graphicx}
\usepackage{dcolumn}
\usepackage{bm}
\usepackage{multirow}
\usepackage{booktabs}
\usepackage[final]{changes}

\begin{document}

\preprint{APS/123-QED}

\title{Rare-earth tuning of magnetic entropy and interaction scales in K$_3$$Ln$(PO$_4$)$_2$ ($Ln$ = Yb, Gd) for adiabatic demagnetization refrigeration}

\author{Yonglin Wang}
\affiliation{School of Physics, Beihang University, Beijing 100191, China}

\author{Chuandi Zhang}
\affiliation{School of Physics, Beihang University, Beijing 100191, China}

\author{Jianglong Zhang}
\affiliation{State Key Laboratory of Quantum Functional Materials, School of Physical Science and Technology, ShanghaiTech University, Shanghai 201210, China}
\affiliation{ShanghaiTech Laboratory for Topological Physics, ShanghaiTech University, Shanghai 201210, China}

\author{Zihang Gao}
\affiliation{School of Physics, Beihang University, Beijing 100191, China}

\author{Ziqi Li}
\affiliation{School of Physics, Beihang University, Beijing 100191, China}

\author{Feiran Shen}
\affiliation{Spallation Neutron Source Science Center, Dongguan 523803, China}
\affiliation{Institute of High Energy Physics, Chinese Academy of Sciences, Beijing 100190, China}

\author{Lunhua He}
\affiliation{Spallation Neutron Source Science Center, Dongguan 523803, China}
\affiliation{Institute of High Energy Physics, Chinese Academy of Sciences, Beijing 100190, China}

\author{Peijie Sun}
\affiliation{Beijing National Laboratory for Condensed Matter Physics, Institute of Physics, Chinese Academy of Sciences, Beijing, China}

\author{Hao Deng}
 \email{denghao@shanghaitech.edu.cn}
\affiliation{State Key Laboratory of Quantum Functional Materials, School of Physical Science and Technology, ShanghaiTech University, Shanghai 201210, China}
\affiliation{ShanghaiTech Laboratory for Topological Physics, ShanghaiTech University, Shanghai 201210, China}

\author{Junsen Xiang}
 \email{xiangjs@iphy.ac.cn}
\affiliation{Beijing National Laboratory for Condensed Matter Physics, Institute of Physics, Chinese Academy of Sciences, Beijing, China}

\author{Wentao Jin}
 \email{wtjin@buaa.edu.cn}
\affiliation{School of Physics, Beihang University, Beijing 100191, China}


\begin{abstract}
Adiabatic demagnetization refrigeration (ADR) based on frustrated rare-earth magnets provides a promising route toward helium-free sub-Kelvin cooling, while balancing magnetic entropy and interaction scales is essential for optimizing the ADR functionalities. Here we investigate the low-temperature magnetic and ADR properties of the alkali rare-earth double phosphates K$_3$$Ln$(PO$_4$)$_2$ ($Ln$ = Yb, Gd), which crystallize in a monoclinic $P2_1/m$ structure containing distorted isosceles triangular lattices of rare-earth ions. Magnetization and specific heat measurements confirm the absence of long-range magnetic ordering down to 0.1~K in both compounds. Direct quasi-ADR measurements demonstrate strongly contrasting refrigeration behaviors: K$_3$Yb(PO$_4$)$_2$ reaches a minimum temperature of 29.8~mK from an initial state of 2~K and 6~T, whereas K$_3$Gd(PO$_4$)$_2$ exhibits extended hold times below 150~mK associated with its larger magnetic entropy reservoir. Crystal-electric-field calculations confirm a $J_{\mathrm{eff}} = 1/2$ ground-state doublet in K$_3$Yb(PO$_4$)$_2$, while thermodynamic analysis based on magnetic entropy and interaction scales accounts for the distinct ADR performance of the two compounds. These results identify K$_3$$Ln$(PO$_4$)$_2$ as a chemically robust platform for rare-earth-tunable solid-state ADR.
\end{abstract}
\maketitle


\section{Introduction}

Achieving temperatures in the sub-Kelvin regime is essential for exploring emergent quantum phenomena, including superconductivity, superfluidity, quantum spin liquids, and topological quantum phases of matter \cite{onnes1911superconductivity,kapitza1938viscosity,wheatley1975experimental,Balents2010,broholm_quantum_2020,tsui1982,wang2017topological}. Although dilution refrigeration remains the dominant technology for ultra-low-temperature cooling, the increasing cost and scarcity of helium resources have stimulated growing interest in alternative cryogenic approaches \cite{siddhantakar2023helium}. Among them, adiabatic demagnetization refrigeration (ADR) provides a versatile solid-state route toward sub-Kelvin cooling by exploiting entropy transfer during isentropic demagnetization of magnetic materials \cite{debye1926einige, giauque1927thermodynamic,weiss1917phenomene}.

Traditional ADR materials are primarily based on hydrated paramagnetic salts such as Mn(NH$_4$)$_2$(SO$_4$)$_2\cdot6$H$_2$O (MAS), NH$_4$Fe(SO$_4$)$_2\cdot12$H$_2$O (FAA), Ce$_2$Mg$_3$(NO$_3$)$_{12}\cdot24$H$_2$O (CMN), and  KCr(SO$_4$)$_2\cdot12$H$_2$O (CPA) \cite{vilches1966measurements, daniels1954thermal, fisher1973magnetothermodynamics}. Their weakly interacting and dilute magnetic moments enable efficient cooling to ultra-low temperatures, but their practical applications are limited by the tendency to dehydration, low volumetric spin density, poor thermal conductivity, and incompatibility with ultra-high-vacuum environments \cite{Zheng2016, wikus2014magnetocaloric}. In addition, their synthesis via solution growth further restricts the scalability and reproducibility.

Recent efforts have therefore shifted toward frustrated quantum magnets, where competing interactions suppress conventional magnetic ordering and preserve substantial magnetic entropy down to very low temperatures. In contrast to dilute paramagnets, these systems combine appreciable spin density with strongly suppressed ordering temperatures, making them promising candidates for ADR \cite{liu2022quantum,tokiwa2021frustrated,treu2024utilizing,li-2024-nc}. Representative examples include the recently discovered triangular-lattice spin supersolids Na$_2$BaCo(PO$_4$)$_2$ and EuCo$_2$Al$_9$ \cite{xiang2024giant,Zhang_2024,Gao_2024,shu2026giant}, where frustration and enhanced low-energy spin fluctuations give rise to pronounced magnetocaloric responses.

Optimizing ADR performance in such systems requires balancing two competing energy scales. On the one hand, the magnetic entropy associated with the crystal-electric-field (CEF) ground-state manifold determines the available entropy reservoir for cooling. On the other hand, the characteristic interaction scale, including exchange and dipolar interactions, governs the development of collective correlations and ultimately limits the minimum attainable temperature. Although these energy scales are intrinsically coupled in real materials, their interplay provides a useful framework for understanding and comparing ADR performance across different magnetic systems.

Rare-earth ions provide a particularly versatile platform for tuning the balance between magnetic entropy and interaction scales. Their localized 4$f$ electrons support diverse CEF ground states, enabling systematic control of the low-temperature magnetic degrees of freedom. In particular, Yb$^{3+}$ often realizes an isolated Kramers doublet described by an effective $J_{\mathrm{eff}} = 1/2$ degree of freedom, with weak interactions and a relatively small entropy reservoir that favors access to ultra-low temperatures. In contrast, Gd$^{3+}$ retains a nearly isotropic $S = 7/2$ multiplet with maximal spin degeneracy, yielding substantially larger magnetic entropy but typically stronger interactions. Recent studies on Yb- and Gd-based borates and phosphates, including $Ln$BO$_3$\cite{mukherjee2018magnetic,lin2026quantum}, KBa$Ln$(BO$_3$)$_2$\cite{tokiwa2021frustrated,PhysRevB107104402}, Ba$_3$$Ln$B$_9$O$_{18}$\cite{klinger2025sub}, Ba$_3$$Ln$B$_3$O$_9$\cite{klinger2025sub}, Na$Ln$P$_2$O$_7$\cite{PhysRevB111064431,PhysRevApplied20014013}, further demonstrate how rare-earth selection governs the balance between attainable temperature and available magnetic entropy in ADR materials.

Within this context, the alkali rare-earth double phosphates $A_3$$Ln$(PO$_4$)$_2$ ($A$ = alkali metal, $Ln$ = rare earth) provide an attractive layered platform for investigating rare-earth-dependent magnetic and ADR properties. Previous studies have established the crystal structures and elementary magnetic properties of this family \cite{farmer2016structural,guo2020crystal,tisdale2022investigation,wang2024large}. For example, Rb$_3$Yb(PO$_4$)$_2$ was identified as an effective spin-1/2 triangular-lattice magnet with suppressed magnetic ordering down to the millikelvin regime \cite{guo2020crystal}, while K$_3$Gd(PO$_4$)$_2$ was reported to exhibit an anomalous low-temperature magnetic transition and magnetocaloric response \cite{wang2024large}. However, how rare-earth-dependent low-energy magnetic manifolds and interaction scales determine the resulting ADR functionalities within this structural family remains largely unexplored.

Here we investigate the low-temperature magnetic and ADR properties of K$_3$$Ln$(PO$_4$)$_2$ ($Ln$ = Yb, Gd), which crystallize in a monoclinic structure containing distorted isosceles triangular lattices of rare-earth ions \cite{farmer2016structural}. Direct quasi-ADR measurements reveal contrasting refrigeration behaviors within the same structural framework. K$_3$Yb(PO$_4$)$_2$ functions as an efficient ultra-low-temperature refrigerant, reaching a minimum temperature of 29.8~mK, whereas K$_3$Gd(PO$_4$)$_2$ serves as a high-entropy refrigerant with extended hold time below 150~mK. The contrasting ADR behaviors highlight the complementary refrigeration functionalities enabled by rare-earth selection within the same structural platform. More broadly, these results demonstrate how rare-earth-dependent low-energy manifolds govern the balance between attainable temperature and entropy reservoir, establishing K$_3$$Ln$(PO$_4$)$_2$ as a promising system for multi-stage solid-state ADR.

\section{Methods}

Polycrystalline K$_3$$Ln$(PO$_4$)$_2$ ($Ln$ = Yb, Gd) samples were synthesized via conventional solid-state reaction using high-purity K$_2$CO$_3$ (99.99\%), NH$_4$H$_2$PO$_4$ (99.99\%), and $Ln_2$O$_3$ (99.99\%, $Ln$ = Yb, Gd). The starting materials were mixed in a molar ratio of K$_2$CO$_3$ : NH$_4$H$_2$PO$_4$ : $Ln_2$O$_3$ = 1.605 : 2 : 0.5, corresponding to a 7\% excess of potassium to compensate for volatilization during the heat treatment. After thorough grinding in an argon-filled glovebox, the mixtures were pelletized and calcined in air at 980~$^\circ$C for 6~h for the Yb compound and 880~$^\circ$C for the Gd compound. Several intermediate grinding and sintering cycles were performed to improve the phase purity.
Single-crystal samples were grown using a flux method with K$_4$P$_2$O$_7$ as the solvent \cite{ushakov2004thermochemistry}. A mixture containing 12~wt\% $Ln_2$O$_3$ and the solvent was heated to 1235~$^\circ$C in a platinum crucible, held for 4~h, and subsequently cooled to 900~$^\circ$C at a controlled rate of 1~$^\circ$C/h.

Room-temperature powder X-ray diffraction (XRD) measurements were performed on a Bruker D8 ADVANCE diffractometer with Cu K$\alpha$ radiation ($\lambda = 1.5406$~\AA) in Bragg-Brentano geometry. Room-temperature neutron powder diffraction (NPD) experiments on K$_3$Yb(PO$_4$)$_2$ were carried out on the General Purpose Powder Diffractometer (31113.02.CSNS.GPPD) at the China Spallation Neutron Source (CSNS, Dongguan, China) \cite{he2023performance}, while corresponding measurements on K$_3$Gd(PO$_4$)$_2$ were not feasible due to the strong neutron absorption of Gd. Structural refinements to the diffraction patterns were conducted via the Rietveld method using the FULLPROF suite \cite{rodriguez1993recent}.

DC magnetization measurements on the polycrystalline samples were performed using a Quantum Design magnetic property measurement system. Zero-field-cooling magnetization $M(T)$ was measured between 1.8 and 300~K under an applied field of 0.1~T. Isothermal magnetization $M(H)$ curves were measured up to 7~T at 1.8, 5, and 10~K. Specific heat measurements down to $\sim$0.1~K were conducted with single-crystal samples using a Quantum Design physical property measurement system (PPMS) equipped with a dilution refrigerator insert. Polycrystalline samples were used for magnetization measurements as they provide sufficient sample mass for reliable magnetic signals, whereas single crystals were selected for low-temperature specific heat measurements to ensure well-defined sample geometry and thermal coupling.

Quasi-ADR measurements were performed using a homemade setup integrated into the PPMS. Polycrystalline K$_3$$Ln$(PO$_4$)$_2$ samples (3~g) were mixed with silver powders in a 1:1 mass ratio to improve the thermal conductivity and subsequently pressed into pellets. The pelletized samples were further surrounded by Gd$_3$Ga$_5$O$_{12}$ (GGG) crystals, which served as a thermal shield to suppress parasitic heat leaks during demagnetization. The sample temperature was monitored utilizing a field-calibrated RuO$_2$ thermometer attached directly to the pellet surface. Demagnetization measurements were initiated from 2~K under applied fields of 2, 4, and 6~T, followed by field sweeps to zero at a constant rate of 50~Oe/s. This rate was selected as a compromise between minimizing eddy-current heating and keeping the demagnetization time within a practical experimental duration. The effective heat leak of the setup, estimated from the post-demagnetization warming curves together with the measured heat capacity of the sample, is on the order of 20--400~nW over the temperature range of 100--300~mK, depending on the sample and experimental conditions.

\section{Results and discussions}
\subsection{Structural characterizations}

\begin{figure*}
 \includegraphics[width=\textwidth]{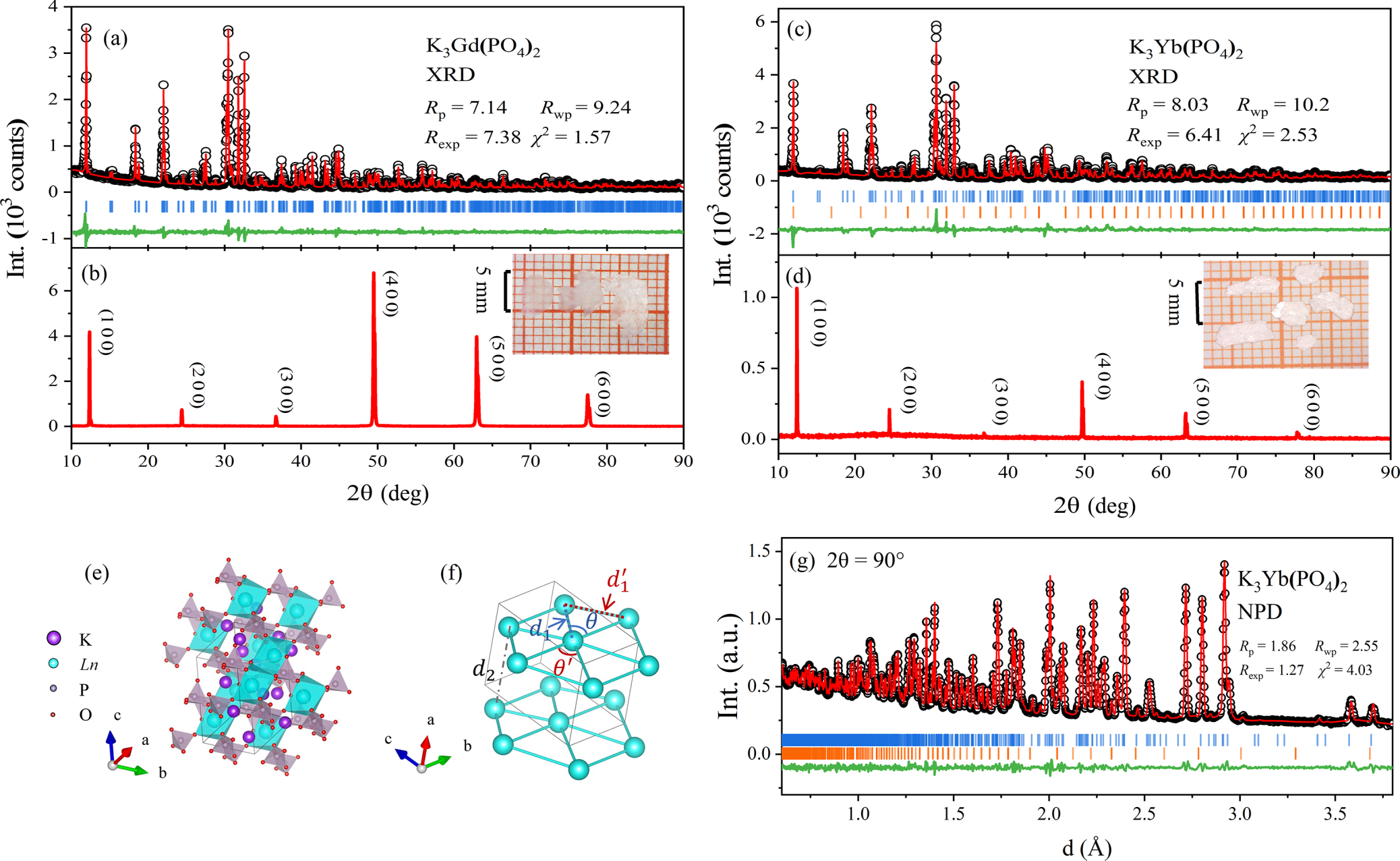}
 \caption{Room-temperature powder XRD refinements of (a) K$_3$Gd(PO$_4$)$_2$ and (c) K$_3$Yb(PO$_4$)$_2$, together with the NPD refinement of (g) K$_3$Yb(PO$_4$)$_2$. Black open circles represent the experimental data, red solid lines denote the calculated patterns, and green solid curves show the differences between the observed and calculated intensities. Blue vertical bars mark the nuclear reflections of K$_3$$Ln$(PO$_4$)$_2$ ($Ln$ = Yb and Gd), while the orange bars (c) and (g) correspond to the impurity phase Yb$_2$O$_3$. (b), (d) Single-crystal XRD patterns of K$_3$Gd(PO$_4$)$_2$ and K$_3$Yb(PO$_4$)$_2$, respectively, with the insets showing images of the corresponding crystals. (e) Crystal structure of K$_3$$Ln$(PO$_4$)$_2$. (f) Schematic illustration of the distorted triangular lattice formed by the $Ln^{3+}$ ions. The in-plane nearest-neighbor distances $d_1$ and $d_1'$, the interplane spacing $d_2$, and the corresponding vertex angles $\theta$ and $\theta'$ are indicated. The distorted triangular layers are stacked along the $a$ axis.}
 \label{XRD}
\end{figure*}

The phase purity of the synthesized samples was examined by powder XRD. As shown in Figs.~\ref{XRD}(a) and \ref{XRD}(c), the K$_3$Gd(PO$_4$)$_2$ sample is a pure phase with no detectable impurities, while the K$_3$Yb(PO$_4$)$_2$ sample contains a minor Yb$_2$O$_3$ impurity of approximately 1 wt.\%. Both compounds crystallize in the monoclinic structure (space group $P2_1/m$). To accurately determine the oxygen positions and refine the rare-earth coordination environment, room-temperature NPD patterns were collected for K$_3$Yb(PO$_4$)$_2$ [see Fig.~\ref{XRD}(g)]. Compared with XRD, NPD provides enhanced sensitivity to oxygen atoms, enabling a more reliable determination of the local crystal geometry relevant to the magnetic properties. The refined structural parameters are summarized in Tables~\ref{tab:table1} and \ref{tab:table2}, consistent with previous reports on these compounds \cite{farmer2016structural}. The lattice of K$_3$Gd(PO$_4$)$_2$ is slightly expanded relative to K$_3$Yb(PO$_4$)$_2$, consistent with the larger ionic radius of Gd$^{3+}$.

The monoclinic structure of K$_3$$Ln$(PO$_4$)$_2$ is illustrated in Fig.~\ref{XRD}(e,f). The $Ln$O$_7$ polyhedra are separated by nonmagnetic PO$_4$ tetrahedra and K$^+$ ions, forming distorted isosceles triangular lattices of $Ln^{3+}$ ions within the $bc$ plane [Fig.~\ref{XRD}(e)]. The structural parameters characterizing the triangular geometry are summarized in Tables~\ref{tab:table1} and \ref{tab:table2}. As shown in Fig.~\ref{XRD}(f), the distortion gives rise to two inequivalent in-plane $Ln$--$Ln$ distances, $d_1$ and $d_1'$, together with two corresponding vertex angles, $\theta$ and $\theta'$. The vertex angles in K$_3$Yb(PO$_4$)$_2$ are closer to the ideal value of 60$^\circ$ than those in K$_3$Gd(PO$_4$)$_2$, indicating a weaker distortion of the triangular lattice in the Yb compound. In both compounds, the interplane spacing $d_2$ is significantly larger than the in-plane $Ln$--$Ln$ distances, suggesting quasi-two-dimensional magnetic lattices with dominant in-plane interactions.

\begin{table}
\caption{\label{tab:table1}Structural parameters of K$_3$Yb(PO$_4$)$_2$ obtained from simultaneous refinements of the room-temperature XRD and NPD patterns.}
\begin{ruledtabular}
\begin{tabular}{c c c c c}
Atom & Site & \textbf{$x$} & \textbf{$y$} & \textbf{$z$} \\
\hline
Yb & 2e & 0.0080(1)   & 0.25      & 0.2858(1)  \\
P1 & 2e & 0.8152(4)   & 0.25      & 0.5727(3)  \\
P2 & 2e & 0.7726(4)   & 0.75      & 0.0916(3)  \\
O1 & 2e & 0.0160(3)   & 0.25      & 0.5338(2)  \\
O2 & 2e & 0.7106(4)   & 0.25      & 0.4294(2)  \\
O3 & 4f & 0.7693(2)   & 0.0258(3) & 0.6618(1)  \\
O4 & 4f & 0.8522(2)   & 0.9745(3) & 0.1679(1)  \\
O5 & 2e & 0.8334(3)   & 0.75      & 0.9339(2)  \\
O6 & 2e & 0.5694(3)   & 0.75      & 0.1020(3)  \\
K1 & 2e & 0.7954(5)   & 0.25      & 0.9201(4)  \\
K2 & 2e & 0.3619(5)   & 0.25      & 0.5845(4)  \\
K3 & 2e & 0.5050(6)   & 0.25      & 0.1922(3)  \\
\hline
\multicolumn{5}{l}{$a = 7.396(1)$~\AA, $b = 5.604(7)$~\AA, $c = 9.315(3)$~\AA,} \\
\multicolumn{5}{l}{$\beta = 91.013(5)^\circ$, $V_{\mathrm{cell}} = 386.08(5)$~\AA$^3$;} \\
\multicolumn{5}{l}{$\theta = 70.0(1)^\circ$, $\theta' = 55.53(3)^\circ$,} \\
\multicolumn{5}{l}{$d_1 = 4.8677(9)$~\AA, $d_1' = 5.5868(15)$~\AA, $d_2 = 7.3619(10)$~\AA.} \\
\end{tabular}
\end{ruledtabular}
\end{table}

\begin{table}
\caption{\label{tab:table2}Structural parameters of K$_3$Gd(PO$_4$)$_2$ obtained from refinements of the room-temperature XRD pattern. }
\begin{ruledtabular}
\begin{tabular}{c c c c c}
Atom & Site & \textbf{$x$} & \textbf{$y$} & \textbf{$z$} \\
\hline
Gd & 2e & 0.0068(3)   & 0.25      & 0.2909(2)  \\
P1 & 2e & 0.8043(1)   & 0.25      & 0.5718(1)  \\
P2 & 2e & 0.7657(1)   & 0.75      & 0.0883(1)  \\
O1 & 2e & 0.0259(2)   & 0.25      & 0.5509(2)  \\
O2 & 2e & 0.7006(2)   & 0.25      & 0.4254(2)  \\
O3 & 4f & 0.7452(1)   & 0.0375(2) & 0.6633(1)  \\
O4 & 4f & 0.8424(1)   & 0.9740(2) & 0.1585(1)  \\
O5 & 2e & 0.8135(3)   & 0.75      & 0.9402(2)  \\
O6 & 2e & 0.5638(2)   & 0.75      & 0.1058(2)  \\
K1 & 2e & 0.7949(8)   & 0.25      & 0.9185(7)  \\
K2 & 2e & 0.3618(8)   & 0.25      & 0.5933(7)  \\
K3 & 2e & 0.5022(1)   & 0.25      & 0.1952(8)  \\
\hline
\multicolumn{5}{l}{$a = 7.4160(1)$~\AA, $b = 5.6254(1)$~\AA, $c = 9.4350(1)$~\AA,} \\
\multicolumn{5}{l}{$\beta = 90.869(1)^\circ$, $V_{\mathrm{cell}} = 393.5698$~\AA$^3$;} \\
\multicolumn{5}{l}{$\theta = 70.9(6)^\circ$, $\theta' = 54.27(5)^\circ$,} \\
\multicolumn{5}{l}{$d_1 = 4.846(4)$~\AA, $d_1' = 5.62544(0)$~\AA, $d_2 = 7.41601(0)$~\AA.} \\
\end{tabular}
\end{ruledtabular}
\end{table}

\begin{table}
\caption{\label{tab:phonon}Fitted Debye and Einstein characteristic temperatures obtained from the Debye-Einstein fits to the phonon specific heat of K$_3$Yb(PO$_4$)$_2$ and K$_3$Gd(PO$_4$)$_2$.}
\begin{ruledtabular}
\begin{tabular}{lcc}
\textbf{Parameter} & \textbf{K$_3$Yb(PO$_4$)$_2$} & \textbf{K$_3$Gd(PO$_4$)$_2$} \\
\hline
$\Theta_D$ (K)  & 159.40 & 143.74 \\
$\Theta_{E1}$ (K) & 125.88 & 149.74 \\
$\Theta_{E2}$ (K) & 223.85 & 230.23 \\
$\Theta_{E3}$ (K) & 492.39 & 495.24 \\
\end{tabular}
\end{ruledtabular}
\end{table}

\subsection{Magnetization}

\begin{figure*}
\includegraphics[width=\textwidth]{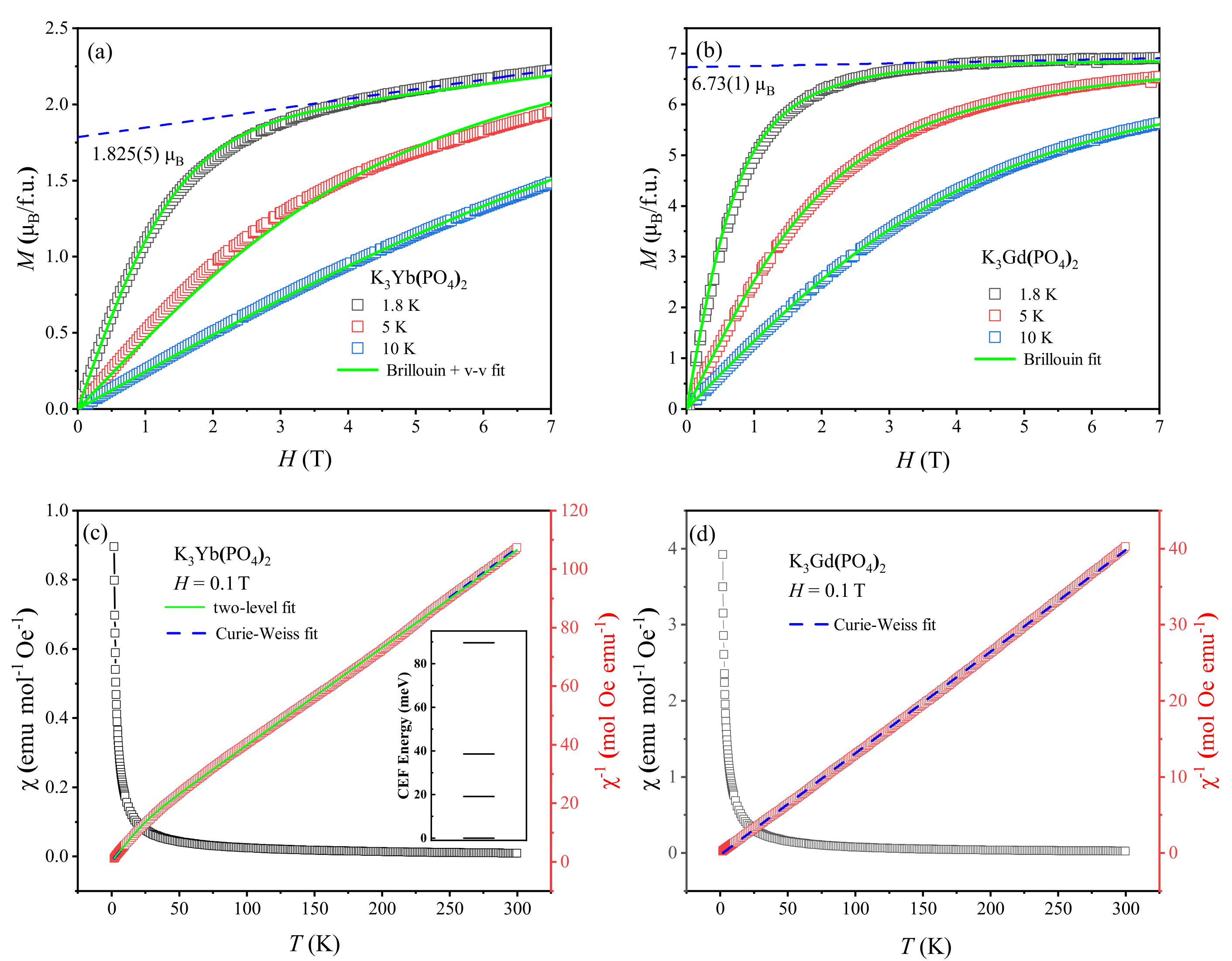}
\caption{Magnetic properties of polycrystalline K$_3$$Ln$(PO$_4$)$_2$ ($Ln$ = Yb and Gd). Isothermal magnetization \(M(H)\) measured at 1.8, 5, and 10 K, for K$_3$Yb(PO$_4$)$_2$ (a) and K$_3$Gd(PO$_4$)$_2$ (b). The dashed lines represent linear fits to the high-field region (5--7~T), while the solid lines correspond to Brillouin-function fits. Temperature dependence of the magnetic susceptibility ($\chi(T)=M/H$) and inverse susceptibility ($\chi^{-1}(T)=H/M$) measured under an applied field of 0.1 T for K$_3$Yb(PO$_4$)$_2$ (c) and K$_3$Gd(PO$_4$)$_2$ (d). The dashed and solid lines denote the CW and phenomenological fits as discussed in the main text. The inset in panel (c) displays the CEF energy levels of K$_3$Yb(PO$_4$)$_2$ calculated from a point-charge model, which reveals a well-isolated ground-state doublet.}
\label{Mag}
\end{figure*}

The temperature- and field-dependent dc magnetization data of the polycrystalline samples are summarized in Fig.~\ref{Mag}. No signatures of long-range magnetic ordering are observed down to 1.8~K in either compound.

For K$_3$Yb(PO$_4$)$_2$, the isothermal magnetization $M(H)$ measured at 1.8~K (Fig.~\ref{Mag}(a)) exhibits a nearly saturated behavior at high fields, together with a linear contribution ($\chi_0 H$) arising from Van Vleck paramagnetism associated with excited CEF levels. Fitting the 5--7~T region using $M = M_{\text{sat}} + \chi_0 H$ yields a saturation moment of $M_{\text{sat}} = 1.825(5)~\mu\rm_B$. The $M(H)$ curves measured at 1.8, 5, and 10~K are all well reproduced by a $J_{\mathrm{eff}} = 1/2$ Brillouin function including the Van Vleck term, yielding an average effective Landé factor of $g \approx 3.48$.

The dc magnetic susceptibility $\chi(T)$ and inverse susceptibility $\chi^{-1}(T)$ of K$_3$Yb(PO$_4$)$_2$ are shown in Fig.~\ref{Mag}(c). $\chi^{-1}(T)$ exhibits two distinct Curie-Weiss (CW) regimes associated with the CEF splitting of the $J = 7/2$ multiplet into four Kramers doublets. The high-temperature (250--300~K) CW fit gives a Weiss temperature $\theta_{\text{H}} = -17.7(5)$~K and an effective moment $\mu_{\text{eff}} = 4.87~\mu\rm_B$, close to the free-ion value of Yb$^{3+}$ ($4.54~\mu\rm_B$). In the low-temperature regime (1.8--10~K), the fitted parameters are $\theta_{\text{L}} = -0.22(1)$~K and $\mu_{\text{eff}} = 3.76~\mu\rm_B$, indicating weak antiferromagnetic interactions. These results further suggest that the well-isolated lowest Kramers doublet with \(J_{\text{eff}} = 1/2\) dominates the low-temperature magnetic response.

To further characterize the crossover between the two CW regimes, the susceptibility was analyzed using a phenomenological two-level model~\cite{besara2014single,singh2024bi}:

\begin{equation}
\chi(T) = \frac{1}{8(T-\theta_{\text{CW}})} \frac{\mu_{1,\text{eff}}^{2} + \mu_{2,\text{eff}}^{2} e^{-\Delta/T}}{1 + e^{-\Delta/T}}.
\label{eq:twolvl}
\end{equation}
Here $\mu_{1,\text{eff}}$ and $\mu_{2,\text{eff}}$ denote the effective moments associated with the ground-state and excited CEF doublets, respectively, $\Delta$ denotes the energy separation between the two levels expressed in temperature units, and $\theta_{\text{CW}}$ represents the effective Weiss temperature. In the fitting procedure, $\theta_{\text{CW}} = -0.22(1)$~K and $\mu_{1,\text{eff}} = 3.76~\mu\rm_B$ were fixed to the values obtained from the low-temperature CW fit. 

The resulting fit (solid line in Fig.~\ref{Mag}(c)) yields $\mu_{2,\text{eff}} = 5.786(4)~\mu\rm_B$ and $\Delta = 80.8(8)$~K. This extracted energy gap is qualitatively consistent with our point-charge model calculations of CEF levels, which show a well-isolated ground-state doublet (see inset of Fig.~\ref{Mag}(c) and the detailed Discussion in Sec.~\ref{sec:ADR}). This sizable CEF splitting ensures that  the $J_{\mathrm{eff}} = 1/2$ description of K$_3$Yb(PO$_4$)$_2$ is robust at low temperatures.

In contrast, the $M(H)$ curves of K$_3$Gd(PO$_4$)$_2$ measured at 1.8~K [Fig.~\ref{Mag}(b)] are well described by a conventional $J = 7/2$ Brillouin function, yielding an effective Landé factor $g = 1.958(1)$ and a saturation moment of $M_{\text{sat}} = 6.73(1)~\mu\rm_B$. The data measured at 5~K and 10~K are equally well reproduced within the same framework, giving consistent $g$ values of 1.960(3) and 1.968(2), respectively, confirming the robustness of the $J = 7/2$ description over the measured temperature range.

As shown in Fig.~\ref{Mag}(d), $\chi^{-1}(T)$ remains nearly linear from 1.8~K to 300~K, consistent with the nearly isotropic $L = 0$ character of Gd$^{3+}$. A CW fit over the entire temperature range results in $\mu_{\text{eff}} = 7.74~\mu\rm_B$ and $\theta = 1.8(1)$~K. Despite the positive Weiss temperature, no signatures of long-range magnetic ordering are detected in the magnetization down to 1.8~K. This is further corroborated by the absence of a $\lambda$-type anomaly in the specific heat down to 0.1~K presented below. The coexistence of relatively weak interactions and suppressed ordering is favorable for maintaining substantial magnetic entropy at low temperatures, which is advantageous for ADR applications.

\subsection{Specific heat}

\begin{figure*}
\includegraphics[width=\textwidth]{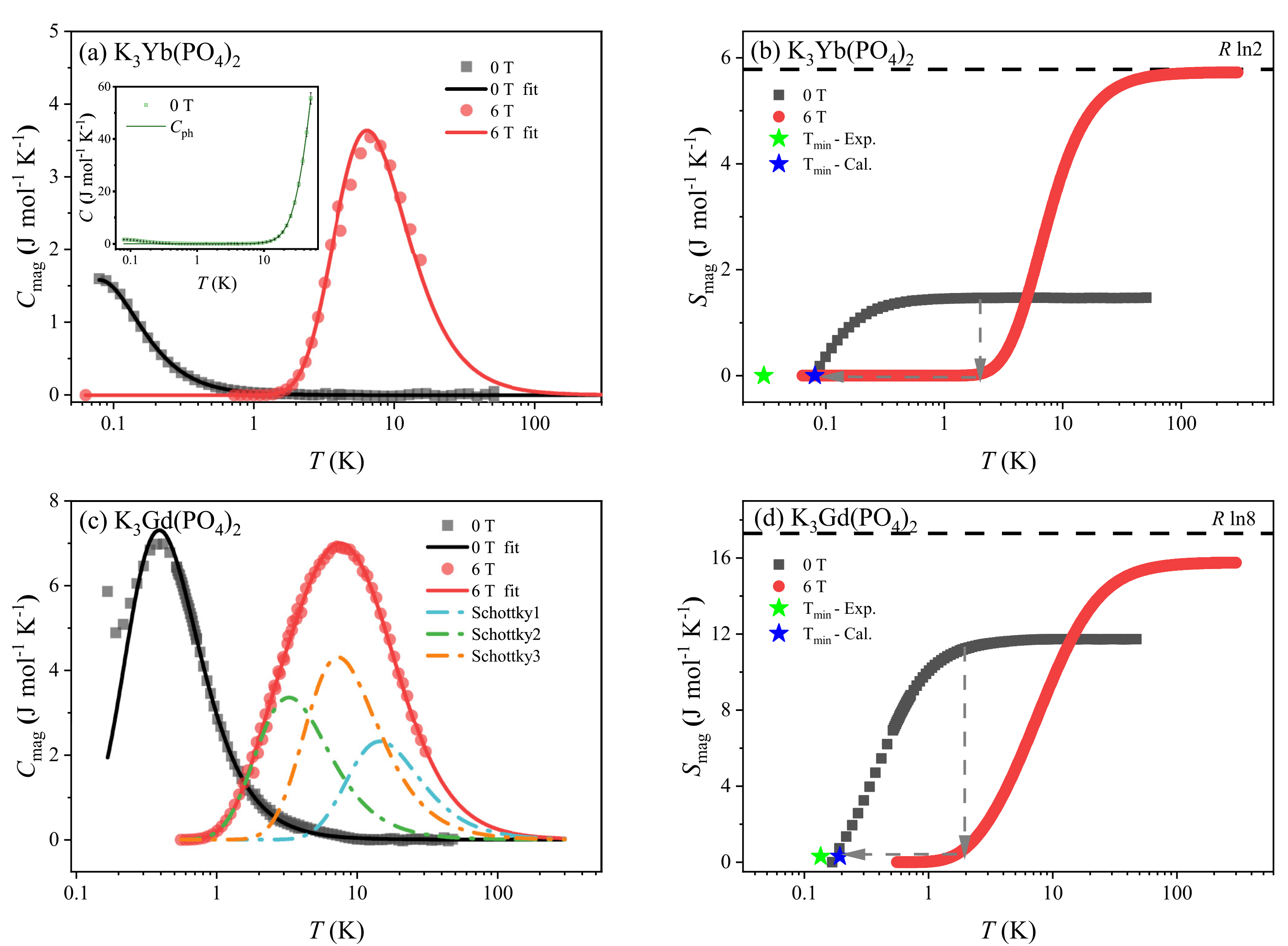}
\caption{Specific heat and magnetic entropy of K$_3$$Ln$(PO$_4$)$_2$ ($Ln$ = Yb and Gd). Magnetic specific heat of K$_3$Yb(PO$_4$)$_2$ (a) and K$_3$Gd(PO$_4$)$_2$ (c) measured under magnetic fields of 0 and 6~T. The inset in panel (a) shows the Debye-Einstein fit to the phonon specific heat of K$_3$Yb(PO$_4$)$_2$. Magnetic entropy ($S_{\mathrm{mag}}$) of K$_3$Yb(PO$_4$)$_2$ (b) and K$_3$Gd(PO$_4$)$_2$ (d) under magnetic fields of 0 and 6~T, obtained by integrating $C_{\mathrm{mag}}(T)/T$. The arrows indicate the ideal adiabatic demagnetization process starting from an initial state of 2~K under 6~T. The blue and green stars mark the calculated and experimentally achieved minimum temperatures, respectively, during the quasi-ADR process.}
\label{C}
\end{figure*}

\begin{figure*}
\includegraphics[width=\textwidth]{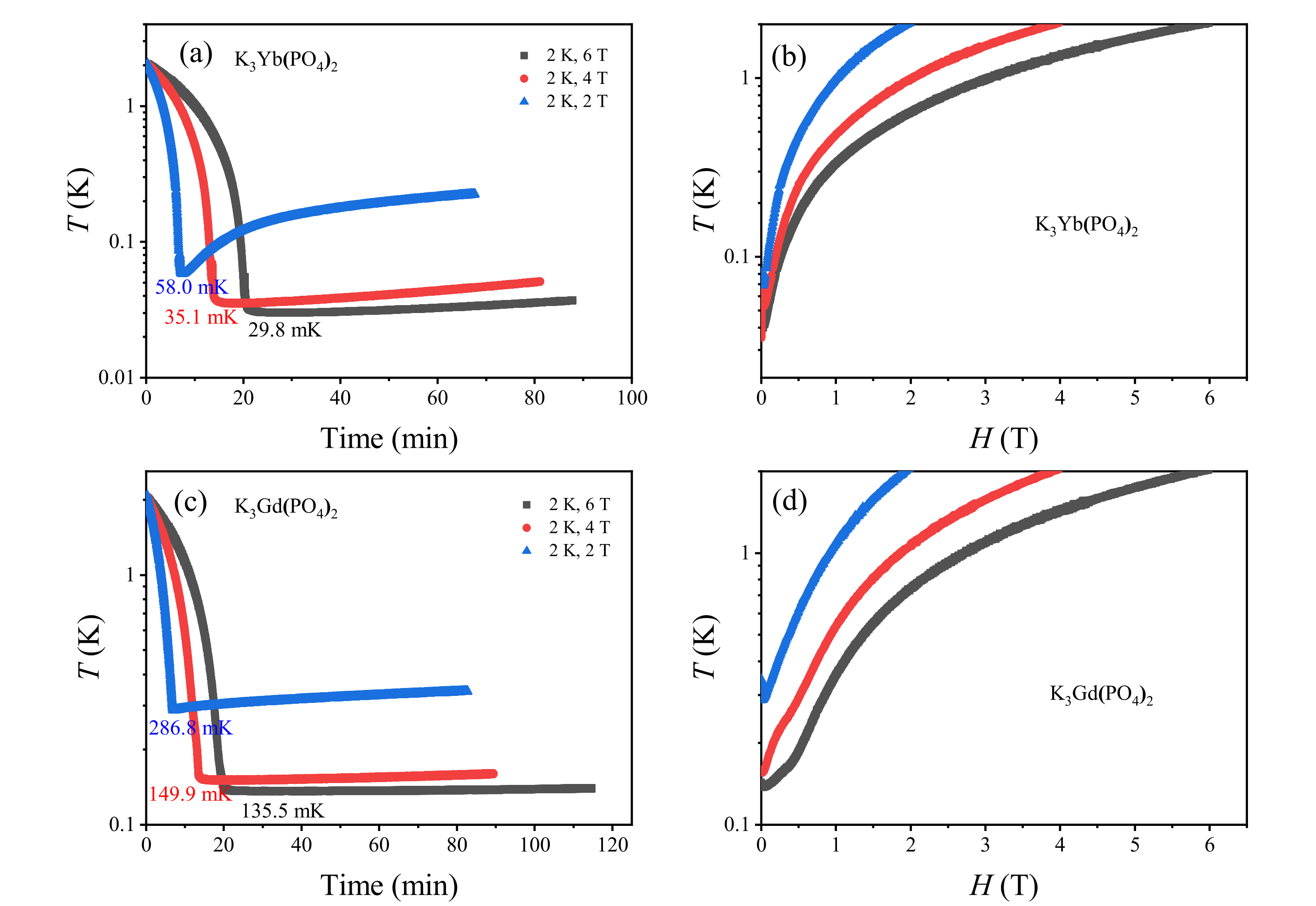}
\caption{ADR performance of K$_3$Yb(PO$_4$)$_2$ (a), (b) and K$_3$Gd(PO$_4$)$_2$ (c), (d), showing the temperature evolutions during the demagnetization process, starting from different initial conditions, as functions of time (a), (c) and magnetic field (b), (d). The black, red, and blue curves correspond to initial magnetic fields of 6, 4, and 2 T, respectively, with all measurements starting from an initial temperature of 2 K.}
\label{ADR}
\end{figure*}

The specific heat of single-crystal samples of K$_3$$Ln$(PO$_4$)$_2$ ($Ln$ = Yb, Gd) provides further insight into their low-energy magnetic properties. As shown in Figs.~\ref{C}(a) and \ref{C}(c), no sharp $\lambda$-type anomaly associated with long-range magnetic ordering was observed down to $\sim$0.1~K in either compound.

The lattice contribution was approximated using a combined Debye-Einstein model consisting of one Debye and three Einstein modes [see inset of Fig.~\ref{C}(a)], where the Debye term accounts for acoustic phonons and the Einstein terms describe the optical phonons: 
\begin{equation}
C_{\mathrm{ph}}(T)=a_D C_D(\Theta_D,T)+\sum_{i=1}^{3}a_{Ei} C_E(\Theta_{Ei},T).
\end{equation}
Here $a_D$ and $a_{Ei}$ denote the weights of the Debye and Einstein contributions, respectively, constrained by the total number of vibrational degrees of freedom,
\begin{equation}
a_D + \sum_{i=1}^{3} a_{Ei} = N = 14,
\end{equation}
with $N = 14$ atoms per formula unit.
The Debye and Einstein contributions are given by
\begin{equation}
C_D(\Theta_D,T)=9R\left(\frac{T}{\Theta_D}\right)^3\int_0^{\Theta_D/T}\frac{x^4 e^x}{(e^x - 1)^2}\, dx,
\end{equation}
\begin{equation}
C_E(\Theta_E,T)=3R\left(\frac{\Theta_E}{T}\right)^2\frac{e^{\Theta_E/T}}{\left(e^{\Theta_E/T} - 1\right)^2},
\end{equation}
where $\Theta_D$ and $\Theta_E$ represent the characteristic Debye and Einstein temperatures, respectively. The fit was performed above 10~K where magnetic contributions become negligible. The fitted parameters summarized in Table~\ref{tab:phonon} are similar for both compounds, indicating comparable phonon energy scales across the series. After subtracting the phonon background, the magnetic specific heat was analyzed using an effective Schottky model, to probe the low-energy magnetic degrees of freedom and the associated entropy evolution.

For K$_3$Yb(PO$_4$)$_2$, the zero-field magnetic specific heat is well described by a single Schottky anomaly with an effective gap of $\Delta = 0.191(17)$~K [Fig.~\ref{C}(a)]. However, the recovered magnetic entropy reaches only about 25\% of $R\ln 2$ within the measured temperature range [Fig.~\ref{C}(b)]. Such incomplete entropy recovery has also been reported in several Yb$^{3+}$-based systems \cite{PhysRevB104L220403,jzd65354,PhysRevB111155141}, suggesting that a fraction of the magnetic entropy may reside below the present experimental temperature range or may be redistributed through low-temperature magnetic correlations. This issue is also relevant to the broader discussions of zero-temperature residual entropy and entropy evolution in geometrically frustrated systems, including spin-ice compounds \cite{ramirez1999zero,pomaranski2013absence}. Under an applied field of 6~T, the Schottky anomaly shifts to $\Delta = 15.43(9)$~K and the full entropy $R\ln 2$ is recovered, consistent with a field-polarized effective $J_{\mathrm{eff}} = 1/2$ ground-state doublet.

In contrast, K$_3$Gd(PO$_4$)$_2$ exhibits no evidence of long-range magnetic ordering despite its positive Weiss temperature. Although a previous magnetization study reported an anomaly near 0.58~K and attributed it to a possible magnetic phase transition \cite{wang2024large}, our single-crystal specific-heat measurements reveal no sharp $\lambda$-type anomaly down to $\sim$0.1~K, suggesting the absence of a bulk thermodynamic transition within the present experimental resolution. This discrepancy may arise from the intrinsic weakness of the low-temperature magnetic anomaly, sample-dependent effects, or the development of short-range magnetic correlations without a transition into a conventional long-range ordered state. Instead, the low-temperature magnetic specific heat is approximately described by a Schottky-like contribution with an effective gap of $\Delta \approx 0.94$~K [Fig.~\ref{C}(c)]. A weak upturn below 0.2~K deviates from the Schottky fit, which may originate from the strengthening of magnetic correlations or a nuclear contribution. Under an applied field of 6~T, the data can no longer be reproduced by a single Schottky term and are instead phenomenologically decomposed into three effective Schottky contributions, consistent with the multi-level Zeeman splitting of the $J = 7/2$ manifold of Gd$^{3+}$. The magnetic entropy approaches the full value of $R\ln 8$ [Fig.~\ref{C}(d)], reflecting the large entropy reservoir associated with the nearly isotropic Gd$^{3+}$ moments.

\begin{table*}
\caption{\label{flo:Eigenvectors} Calculated CEF eigenvalues and eigenvectors for Yb$^{3+}$ in K$_3$Yb(PO$_4$)$_2$ based on a point-charge model~\cite{scheie2021pycrystalfield}. The wave functions are expressed in the $|J, M_J\rangle$ basis with $J = 7/2$.}
\begin{ruledtabular}
\begin{tabular}{c|cccccccc}
$E$ (meV) &$| -\frac{7}{2}\rangle$ & $| -\frac{5}{2}\rangle$ & $| -\frac{3}{2}\rangle$ & $| -\frac{1}{2}\rangle$ & $| \frac{1}{2}\rangle$ & $| \frac{3}{2}\rangle$ & $| \frac{5}{2}\rangle$ & $| \frac{7}{2}\rangle$ \\
 \hline 
0.000 & $-$0.4017 & $-$0.0500 & 0.4115 & 0.4888 & $-$0.4383 & $-$0.3091 & 0.3742 & $-$0.0127 \\
0.000 & 0.0127 & 0.3742 & 0.3091 & $-$0.4383 & $-$0.4888 & 0.4115 & 0.0500 & $-$0.4017 \\
19.107 & 0.5943 & $-$0.4291 & $-$0.1754 & 0.0596 & $-$0.2494 & 0.2020 & 0.5704 & 0.0005 \\
19.107 & $-$0.0005 & 0.5704 & $-$0.2020 & $-$0.2494 & $-$0.0596 & $-$0.1754 & 0.4291 & 0.5943 \\
38.584 & $-$0.0193 & 0.3642 & $-$0.6098 & 0.2785 & $-$0.0118 & $-$0.1930 & 0.1413 & $-$0.6001 \\
38.584 & 0.6001 & 0.1413 & 0.1930 & $-$0.0118 & $-$0.2785 & $-$0.6098 & $-$0.3642 & $-$0.0193 \\
89.589 & 0.3531 & 0.4324 & 0.3906 & 0.5063 & 0.4113 & 0.3192 & 0.0914 & 0.0000 \\
89.589 & 0.0000 & $-$0.0914 & 0.3192 & $-$0.4113 & 0.5063 & $-$0.3906 & 0.4324 & $-$0.3531 \\
\end{tabular}
\end{ruledtabular}
\end{table*}

The magnetic entropy ($S_{\mathrm{mag}}$) was obtained by integrating the magnetic specific heat according to $S_{\mathrm{mag}}(T)=\int_0^T [C_{\mathrm{mag}}(T')/T']\,dT'$, taking $S_{\mathrm{mag}}(0)=0$ as the reference, further providing a thermodynamic framework for evaluating the ADR performance of K$_3$$Ln$(PO$_4$)$_2$. As shown in Figs.~\ref{C}(b) and \ref{C}(d), an ideal demagnetization from an initial state of 2~K under 6~T follows constant-entropy trajectories connecting the finite-field and zero-field entropy curves. From the entropy differences at 2~K, the isothermal magnetic entropy changes ($-\Delta S_M$) induced by 6~T are estimated to be 1.46 and 11.2~J~mol$^{-1}$~K$^{-1}$ for K$_3$Yb(PO$_4$)$_2$ and K$_3$Gd(PO$_4$)$_2$, respectively. These values remain below the ideal limits of $R\ln 2$ and $R\ln 8$. For K$_3$Yb(PO$_4$)$_2$, this reduction is consistent with the incomplete entropy recovery at zero field discussed above. For K$_3$Gd(PO$_4$)$_2$, the reduced entropy change likely reflects the combined effects of low-temperature magnetic correlations and incomplete field polarization arising from thermal population of the Zeeman-split $J=7/2$ levels. Consequently, K$_3$Yb(PO$_4$)$_2$ can access the deep millikelvin regime owing to its reduced interaction scale and strongly suppressed ordering tendency, whereas K$_3$Gd(PO$_4$)$_2$ retains a substantially larger entropy reservoir at the expense of a higher minimum attainable temperature. The experimentally achieved minimum temperatures from direct quasi-ADR measurements, indicated by the green stars in Figs.~\ref{C}(b) and \ref{C}(d), are in reasonable agreement with these thermodynamic estimates (blue stars). Notably, the experimentally achieved minimum temperatures of 29.8~mK and 135.5~mK [green stars in Figs.~\ref{C}(b) and \ref{C}(d)] are substantially lower than the thermodynamically estimated values of 81 and 193~mK [blue stars in Figs.~\ref{C}(b) and \ref{C}(d)], for K$_3$Yb(PO$_4$)$_2$ and K$_3$Gd(PO$_4$)$_2$, respectively. This discrepancy is consistent with the incomplete zero-field entropy recovery discussed above, suggesting that the residual magnetic entropy remaining below the experimental temperature range may lead to an overestimation of the entropy-based minimum temperatures.

\subsection{\label{sec:ADR}ADR performance}

\begin{table*}[htbp]
\caption{\label{tab:ADR parameters}Comparison of magnetic-ion density and volumetric ground-state magnetic entropy of representative ADR materials. $S_{\mathrm{GS}}$ is the ground-state multiplet entropy and $R$ is the universal gas constant. The abbreviations are: MAS = $\mathrm{Mn(NH_4)_2(SO_4)_2\cdot6H_2O}$, FAA = $\mathrm{NH_4Fe(SO_4)_2\cdot12H_2O}$, CPA = $\mathrm{KCr(SO_4)_2\cdot12H_2O}$, CMN = $\mathrm{Mg_3Ce_2(NO_3)_{12}\cdot24H_2O}$.}
\begin{ruledtabular}
\begin{tabular}{l c c c c}
Materials & Mag. ion & Mag. ion density (nm$^{-3}$) & $S_{\mathrm{GS}}$ & $S_{\mathrm{GS}}$/vol. [mJ K$^{-1}$ cm$^{-3}$] \\
\hline
MAS\cite{vilches1966measurements} & Mn$^{2+}$ & 2.8 & $R\ln 6$ & 70 \\
FAA\cite{vilches1966measurements} & Fe$^{3+}$ & 2.1 & $R\ln 6$ & 53 \\
CPA\cite{daniels1954thermal} & Cr$^{3+}$ & 2.2 & $R\ln 4$ & 42 \\
CMN\cite{fisher1973magnetothermodynamics} & Ce$^{3+}$ & 1.7 & $R\ln 2$ & 16 \\
$\mathrm{KBaYb(BO_3)_2}$\cite{tokiwa2021frustrated} & Yb$^{3+}$ & 6.7 & $R\ln 2$ & 64 \\
$\mathrm{NaYbP_2O_7}$\cite{PhysRevApplied20014013} & Yb$^{3+}$ & 6.7 & $R\ln 2$ & 64 \\
$\mathrm{Ba_3YbB_9O_{18}}$\cite{klinger2025sub} & Yb$^{3+}$ & 2.5 & $R\ln 2$ & 24 \\
$\mathrm{K_3Yb(PO_4)_2}$ (this work) & Yb$^{3+}$ & 5.2 & $R\ln 2$ & 50 \\
$\mathrm{KBaGd(BO_3)_2}$\cite{PhysRevB107104402} & Gd$^{3+}$ & 6.7 & $R\ln 8$ & 192 \\
$\mathrm{Ba_3GdB_9O_{18}}$\cite{klinger2025sub} & Gd$^{3+}$ & 2.3 & $R\ln 8$ & 66.8 \\
$\mathrm{K_3Gd(PO_4)_2}$ (this work) & Gd$^{3+}$ & 5.0 & $R\ln 8$ & 144 \\
\end{tabular}
\end{ruledtabular}
\end{table*}

The ADR performance of K$_3$$Ln$(PO$_4$)$_2$ ($Ln$ = Yb, Gd) is summarized in Fig.~\ref{ADR}. Starting from an initial condition of 2~K under 6~T, K$_3$Yb(PO$_4$)$_2$ reaches a minimum temperature ($T\rm_{min}$) of 29.8~mK and remains below 40~mK for approximately 1~h, demonstrating efficient access to the deep millikelvin regime. In contrast, K$_3$Gd(PO$_4$)$_2$ reaches a higher $T\rm_{min}$ of 135.5~mK but maintains temperatures below 150 mK for over 1.5~h, reflecting its substantially larger magnetic entropy reservoir and enhanced thermal stability during demagnetization. These results demonstrate how distinct low-energy magnetic manifolds within the same structural framework give rise to markedly different ADR functionalities.

The temperature evolution during demagnetization is shown in Figs.~\ref{ADR}(b) and \ref{ADR}(d). As the magnetic field is swept to zero, the sample temperature decreases monotonically before reaching a minimum and subsequently rising due to residual heat leaks from the measurement platform. The demagnetization curves remain smooth over most of the field range, indicating efficient thermalization throughout the quasi-ADR process. For K$_3$Yb(PO$_4$)$_2$, the $T_{\mathrm{min}}$ values obtained under initial fields of 6, 4, and 2~T are 29.8, 35.1, and 58.0~mK, respectively, whereas K$_3$Gd(PO$_4$)$_2$ reaches 135.5, 149.9, and 286.8~mK under the same conditions. In both compounds, larger initial magnetic fields lead to lower final temperatures, consistent with the larger entropy changes available during demagnetization. The contrasting ADR behaviors further reflect the balance between interaction scale and accessible magnetic entropy discussed above.

To quantify the low-temperature magnetic energy scales relevant to ADR, we estimate the characteristic dipolar interaction energy $E_D$ using the experimentally measured saturation moments and refined nearest-neighbor distances, following Ref.~\cite{PhysRevB104L220403}. Using $M_{\mathrm{sat}} = 1.825~\mu\rm_B$ and $r_{\text{Yb-Yb}} = 4.8677$~\AA\ for K$_3$Yb(PO$_4$)$_2$, and $M_{\mathrm{sat}} = 6.73~\mu\rm_B$ and $r_{\text{Gd-Gd}} = 4.846$~\AA\ for K$_3$Gd(PO$_4$)$_2$, the estimated dipolar scales $E_D/k\rm_B$ ($\propto \mu_{\text{sat}}^2 / r^3$) are approximately 18.1 and 248~mK, respectively.

For K$_3$Yb(PO$_4$)$_2$, the estimated dipolar scale is comparable to the experimentally achieved $T_{\mathrm{min}}$ of 29.8~mK, suggesting that the ADR performance approaches an interaction-limited regime. In contrast, the Weiss temperature of K$_3$Gd(PO$_4$)$_2$ ($\theta = 1.8$~K) substantially exceeds the estimated dipolar scale, indicating that additional exchange interactions, likely mediated through extended $Ln$-O-P-O-$Ln$ super-superexchange pathways, play a nonnegligible role in the low-temperature magnetic behavior. The correspondingly larger interaction scale naturally limits access to lower temperatures in the Gd compound despite its larger magnetic entropy reservoir.

The effective $J_{\text{eff}} = 1/2$ behavior of K$_3$Yb(PO$_4$)$_2$ is further supported by CEF level calculations with a point-charge model using the package PYCRYSTALFIELD \cite{scheie2021pycrystalfield}, as visually demonstrated in the inset of Fig.~\ref{Mag}(c). As summarized in Table~\ref{flo:Eigenvectors}, the $J = 7/2$ multiplet of Yb$^{3+}$ splits into four Kramers doublets, with the first excited state separated by a large gap of $\Delta_{\mathrm{CEF}} = 19.107$~meV ($\sim 221.7$~K). This substantial energy separation not only explains the crossover in susceptibility but also confines the low-temperature magnetic entropy to the ground-state doublet throughout the ADR temperature range, thereby stabilizing the $J_{\text{eff}} = 1/2$ manifold.

Finally, comparison with conventional hydrated ADR salts and recently developed anhydrous rare-earth refrigerants (Table~\ref{tab:ADR parameters}) highlights the advantages of the K$_3$$Ln$(PO$_4$)$_2$ family in terms of volumetric entropy density and chemical robustness. Traditional hydrated salts such as FAA, CPA, and CMN remain important ADR materials, but their applications in compact devices are limited by water-related chemical instability and relatively low volumetric entropy density. In contrast, the anhydrous phosphate framework of K$_3$$Ln$(PO$_4$)$_2$ enables dense packing of magnetic ions while maintaining the chemical stability. For $J_{\mathrm{eff}}=1/2$ refrigerants targeting the deep millikelvin regime, K$_3$Yb(PO$_4$)$_2$ exhibits a magnetic-ion density of 5.2~nm$^{-3}$ and a corresponding volumetric entropy of 50~mJ/(K~cm$^3$), substantially exceeding that of CMN and Ba$_3$YbB$_9$O$_{18}$~\cite{fisher1973magnetothermodynamics,klinger2025sub} and approaching those of other high-density Yb-based refrigerants such as KBaYb(BO$_3$)$_2$ and NaYbP$_2$O$_7$~\cite{tokiwa2021frustrated,PhysRevApplied20014013}. The relatively moderate magnetic-ion density compared with the latter compounds may also help suppress dipolar interactions, which is favorable for accessing the sub-30~mK regime. Similarly, as large-spin refrigerants, K$_3$Gd(PO$_4$)$_2$ combines the large $R\ln 8$ entropy of the Gd$^{3+}$ ions with a high magnetic-ion density (5.0~nm$^{-3}$), yielding a volumetric entropy density of 144~mJ/(K~cm$^3$), substantially exceeding those of traditional hydrated salts (MAS and FAA)~\cite{vilches1966measurements} as well as Ba$_3$GdB$_9$O$_{18}$~\cite{klinger2025sub}, while approaching the high-density Gd-based refrigerant KBaGd(BO$_3$)$_2$~\cite{PhysRevB107104402}. Together with their distinct ADR behaviors, these results demonstrate that the K$_3$$Ln$(PO$_4$)$_2$ family provides a versatile rare-earth-tunable platform for helium-free solid-state refrigeration. In particular, K$_3$Gd(PO$_4$)$_2$ is well suited for high-entropy precooling stages in multi-stage ADR systems, and K$_3$Yb(PO$_4$)$_2$ enables efficient access to the deep millikelvin regime below 30~mK.

\section{Conclusion}
In summary, we have investigated the low-temperature magnetic and ADR properties of the rare-earth double phosphates K$_3$$Ln$(PO$_4$)$_2$ ($Ln$ = Yb, Gd). Despite sharing the same structural framework composed of distorted triangular rare-earth lattices, the two compounds exhibit markedly different refrigeration behaviors governed by their distinct low-energy magnetic manifolds. K$_3$Yb(PO$_4$)$_2$ realizes a $J_{\mathrm{eff}} = 1/2$ ground-state doublet with strongly suppressed interaction scales, enabling access to the deep millikelvin regime with a minimum temperature of 29.8~mK. In contrast, K$_3$Gd(PO$_4$)$_2$ retains the large entropy reservoir of the nearly isotropic $J = 7/2$ manifold, leading to enhanced thermal stability. By combining thermodynamic analysis, direct quasi-ADR measurements, and CEF calculations, our results establish a clear correlation between the attainable cooling temperature, magnetic entropy, and interaction scale in rare-earth ADR materials. More broadly, this work demonstrates how rare-earth selection provides an effective route for engineering complementary refrigeration functionalities within a common structural platform. The chemically robust and anhydrous nature of the K$_3$$Ln$(PO$_4$)$_2$ family further highlights its potential for helium-free solid-state refrigeration and multi-stage ADR applications.

\begin{acknowledgments}
We acknowledge the financial support from the National Key Projects for Research and Development of China (Grant No. 2023YFA1406003), the National Natural Science Foundation of China (Grants No. 12574154, No. 12074024, and No. 12404180), the Fundamental Research Funds for the Central Universities in China, and the Large Scientific Facility Open Subject of Songshan Lake (Dongguan, Guangdong, Grant No. KFKT2022B05). This work is also supported by the Synergetic Extreme Condition User Facility (SECUF)~\cite{secuf2026}. We thank the staff members of the General Purpose Powder Diffractometer~\cite{gppd2026} at the China Spallation Neutron Source (CSNS) for providing technical support and assistance in data collection and analysis.
\end{acknowledgments}

\section*{Data Availability}
The data that support the findings of this article are publicly available~\cite{wang2026klpozenodo}.

\bibliography{reference}

\end{document}